\documentclass[twocolumn,showpacs,floats,aps,prl,reprint,groupedaddress,amsmath,amssymb]{revtex4-1}
\usepackage{graphicx}
\usepackage{subfigure}
\usepackage{hyperref} 
\usepackage{bm}
\usepackage{float}
\usepackage{yfonts}
\usepackage[titletoc,toc,title]{appendix}

\begin{document}

\title{Andreev bound states formation and quasiparticle trapping in quench dynamics revealed by 
time-dependent counting statistics}
\author{R. Seoane Souto, A. Mart\'{\i}n-Rodero and A. Levy Yeyati}
\affiliation{Departamento de F\'{i}sica Te\'{o}rica de la Materia Condensada,\\
Condensed Matter Physics Center (IFIMAC) and Instituto Nicol\'{a}s Cabrera,
Universidad Aut\'{o}noma de Madrid E-28049 Madrid, Spain}
\date{\today}

\begin{abstract}
We analyze the quantum quench dynamics in the formation of a phase-biased superconducting nanojunction. We find that in the absence 
of an external relaxation mechanism and for very general conditions the system gets trapped in a metastable state, corresponding 
to a non-equilibrium population of the Andreev bound states. The use of the time-dependent Full Counting Statistics (FCS) analysis 
allows us to extract information on the asymptotic population of even and odd many-body states, demonstrating that a {\it universal} 
behavior, dependent only on the Andreev states energy, is reached in the quantum point contact limit. These results shed light on recent experimental observations 
on quasiparticle trapping in superconducting atomic contacts. 
\end{abstract}

\maketitle

{\it Introduction:}
Superconducting nanodevices are of central interest as building blocks of
future quantum information processors. While traditionally based on Josephson junctions architectures
\cite{devoret}, superconducting quantum dots and quantum point
contacts are now being explored for quantum information applications like 
generation of electron entanglement through Cooper pair splitting \cite{hofstetter,hermann}
and the so-called Andreev qubits \cite{landry}. Similar hybrid systems can host Majorana-like 
excitations whose search and potential applications are generating a
huge research activity in the condensed matter community \cite{review-majorana}. 
Within this context, studies of the transient response 
of superconducting nanodevices are of basic as well as practical interest \cite{previous-transient}. 
This is connected with the unexplained evidence of residual non-equilibrium 
quasiparticles which undermines the quantum coherence in these devices \cite{various-qp}. 
In the case of superconducting atomic contacts (SACs) this phenomenon 
manifests in the presence of long lived trapped quasiparticle states 
within their Andreev Bound States (ABS)  \cite{zigrist,theory-poisoning}.

\begin{figure}
\includegraphics[width=1\linewidth]{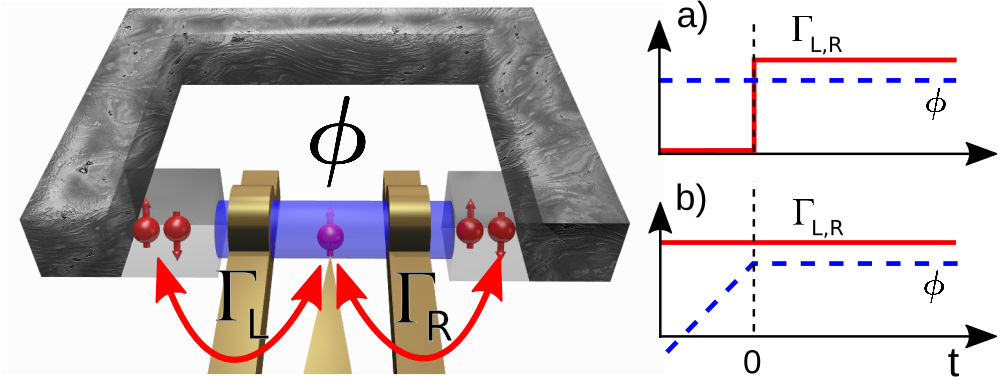}
\caption{(Color online): Schematic representation of the nanojunction formation process considered in this work. In case a)  
the tunneling amplitudes between the central region and the leads are suddenly connected at $t=0$, while in case b) the tunnel amplitudes are constant but 
there is a bias voltage switch off at $t=0$. The nanojunction is formed on a 
superconducting loop threaded by a magnetic field which allows to fix the phase difference $\phi$ .}
\label{fig1}
\end{figure}

In the present work we address these questions by analyzing the quench
dynamics in the formation of a phase-biased superconducting single channel 
contact. We consider the situation schematically depicted in Fig. \ref{fig1}, where
a central electronic level is abruptly coupled to two 
superconducting leads (Fig. \ref{fig1}a). 
Our main question concerns the properties of the state which is generated
at intermediate times (i.e. $\tau_{in} \gg t \gg \hbar/\Delta$, where
$\tau_{in}$ is a characteristic inelastic relaxation time and $\Delta$
is the superconducting gap). We find that for generic values of the parameters
the system gets trapped into a {\it metastable} state, reflecting a non-equilibrium population of the ABSs and exhibiting a smaller
or even opposite supercurrent to the one expected for thermal equilibrium.
While for weak coupling to the leads $\Gamma < \Delta$ this state depends 
strongly on the initial conditions, at large $\Gamma$ it reaches a 
``universal" behavior only dependent on the Andreev level position within
the gap, but still deviating from the equilibrium population.
Furthermore, we study the transient process in terms of charge transfer
probabilities derived from the time-dependent Full Counting Statistics
(FCS) analysis. This allows us to determine the separate populations of
even and odd parity states, an information which is inaccessible from 
any mean field study of single particle properties. We find that the odd parity states,
corresponding to the trapping of a quasiparticle within the ABSs get a
significant population $\sim 0.2-0.5$ for a broad range of parameters, in agreement with the experimental observations for SACs \cite{landry,zigrist}.

{\it Model and formalism:} Our model nanojunction is composed of three regions;
a central spin-degenerate electron level (which can be initially empty or occupied) and two BCS superconducting leads. These are connected at $t=0$ by a tunnel Hamiltonian 
$H_T(t)$, creating excited quasiparticles which undergo multiple tunneling events between the electrodes. In particular, 
successive Andreev reflections are needed to develop the ABSs within the gap at energies $\pm \epsilon_A$ and to establish a non-dissipative (Josephson) current
through the device. The system 
Hamiltonian $H=H_{leads}+H_{0}+H_T$, can be written in terms of Nambu spinors 
$\hat{\Psi}_j=\left(c^{\dagger}_{j\uparrow},c_{j\downarrow}\right)$, where 
$j=k\nu,0$ denotes the $\nu=L,R$ lead and the central level respectively. 
We have $H_{0}=\hat{\Psi}^{\dagger}_{0}\hat{h}_{0}\hat{\Psi}_{0}$, 
$H_{leads}=\sum_{k\nu}\hat{\Psi}^{\dagger}_{k\nu}\hat{h}_{k\nu}\hat{\Psi}_{k\nu}$ and $H_T=\sum_{k,\nu}\left[\hat{\Psi}^{\dagger}_{k\nu}\hat{V}_{\nu}(t)\hat{\Psi}_{0}+\mbox{h.c.}\right]$, where 
$\hat{h}_{0}=\epsilon_0\sigma_z$, $\hat{h}_{k\nu}= \epsilon_{k\nu}\sigma_z + \Delta_{\nu}\sigma_x$ ($\sigma_z$ and $\sigma_x$ denote here Pauli matrices in Nambu space). 
For describing an abrupt switch-on into a phase-biased situation
we use $\hat{V}_{\nu}(t) = \theta(t) V^0_{\nu}\sigma_z e^{i\sigma_z\phi_{\nu}}$,
where $\phi_{L}-\phi_{R} = \phi$ determines the phase difference between the
leads and $\theta(t)$ is the Heaviside function. While this initialization can be considered somewhat artificial, it generates the same dynamics 
as a large applied bias voltage which is suddenly switched off at $t=0$ (see Fig. 1b and SM \cite{supplementary}). For simplicity we consider a constant normal 
density of states $\rho_{L,R}$ in the leads with a finite bandwidth $W$ taken as the larger energy scale in the model and we
define the stationary tunneling rates as $\Gamma_{\nu} = \pi (V^0_{\nu})^2 \rho_{\nu}$, and $\Gamma=\Gamma_L+\Gamma_R$. 
The central level initial charge will be denoted by $n_{\sigma}(0)$, where $\sigma\equiv\uparrow,\downarrow$. Hereafter we assume $\hbar=e=1$.

The transport properties of the system are fully characterized by the 
generating function (GF) defined on the Keldysh contour as \cite{levitov}
\begin{equation}
 Z(\chi,t)= \left\langle T_K exp\left[-i\int_C dt' H_{T,\chi}(t')\right]\right\rangle_0,
 \label{GF}
\end{equation}
where $\chi \equiv \chi_{\nu}(t)$ are counting fields entering as phase factors 
modulating the hopping terms in $H_T$ and
having opposite values $\pm\chi_{\nu}$ on the two branches of the Keldysh 
contour. The average in Eq. (\ref{GF}) is taken over the decoupled system. 
The GF gives access to the charge transfer cumulants, i.e. 
$C_n(t) = (i)^n \partial^n S/\partial \chi^n\rfloor_0$, 
where $S = \ln Z(\chi,t)$. 
For definiteness we will hereafter assume $\chi_L=\chi$ and $\chi_R=0$,
thus focussing on charge transfer through the left interface. The 
corresponding current cumulants are given by 
$I^n(t) = \partial C_n/\partial t$.
One can also decompose the GF as $Z(\chi,t) = \sum P_n(t) e^{i\chi n}$, where $P_n(t)$ can be associated with 
the probability of transferring $n$ charges in the measuring time $t$.
In the BCS superconducting case, the charge in the leads 
is not well defined, and $P_n(t)$ can eventually take negative values
\cite{belzig,ramer,clerk}. The $P_n(t)$ are therefore referred to as 
{\it quasi-probabilities}. 
  
It can be shown that $Z(\chi,t)$ can
be computed as a Fredholm determinant on the Keldysh contour \cite{mukamel,chinos,us}. A straightforward extension of this formalism to 
the superconducting case leads to
\begin{equation}
 Z(\chi,t)=\mbox{det}\left[{\bf G}(\chi=0){\bf G}(\chi)^{-1} \right],
 \label{fredholm}
\end{equation}
 where ${\bf G} = -i \left\langle T_K \Psi_{0}(t) \Psi_0^{\dagger}(t') \right\rangle$
 is the Green function of the dot coupled to the leads defined in Keldysh-Nambu space.
For a generic situation we evaluate Eq. (\ref{fredholm}) numerically following the
approach described in the SM \cite{supplementary}. 
Analytical results allow us to further clarify our findings in certain limits as described below. 

\begin{figure}
\includegraphics[width=1\linewidth]{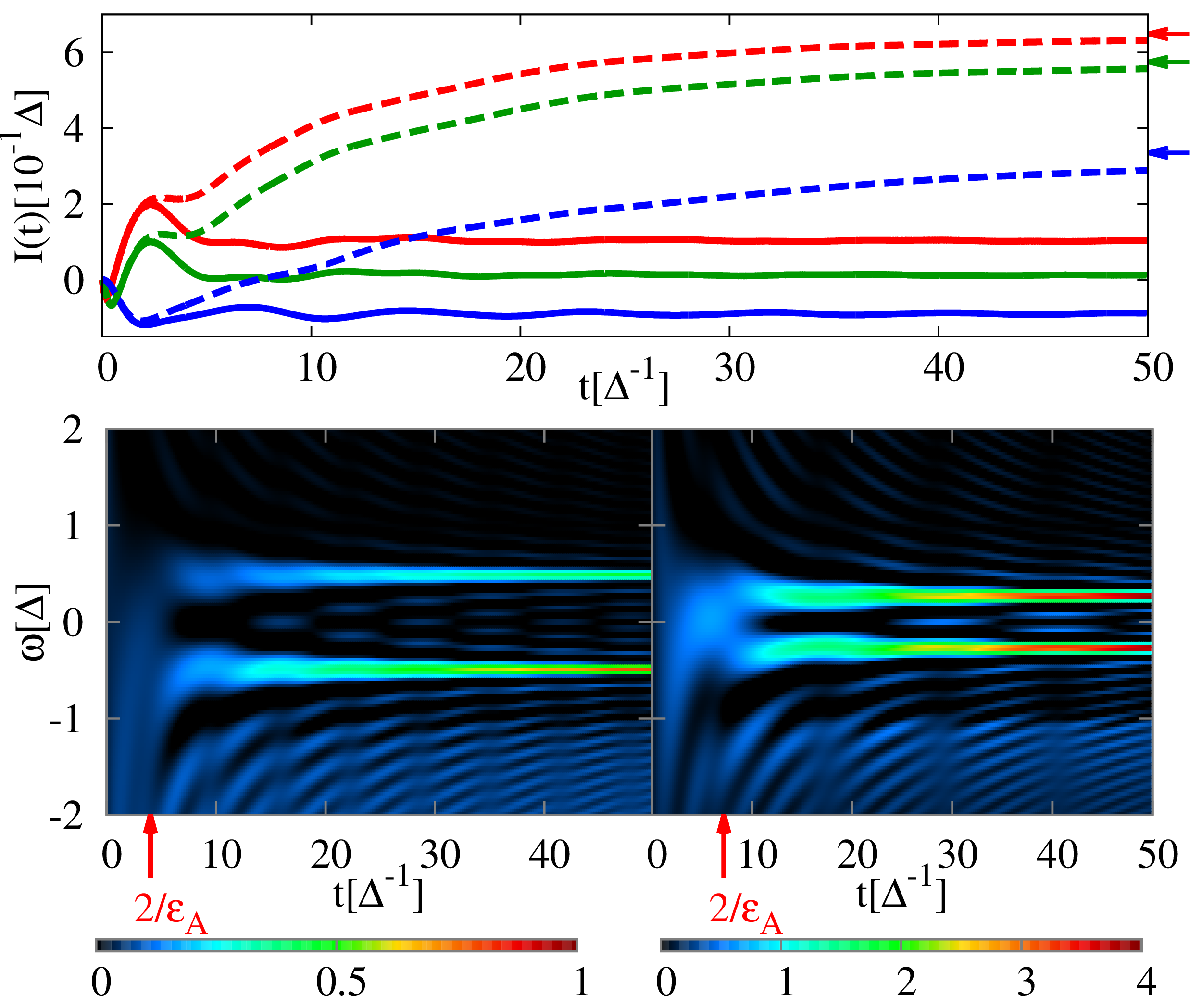}
\caption{(Color online): The upper panel shows the transient current for different 
$\Gamma/\Delta$ values (from top to bottom: 10, 5 and 1) and a fixed phase
difference $\phi/\pi=0.64$ for the perfect transmission case. The full lines 
correspond to the case of no external relaxation mechanism and the dashed ones
correspond to adding a phenomenological broadening in the spectral densities
of size $1/\tau_{in} = 0.02 \Delta$ \cite{comment}. The arrows indicate the 
stationary values for thermal equilibrium. The lower panels show the 
corresponding time-dependent occupied spectral densities for $\Gamma/\Delta$=10 (left) and $\Gamma/\Delta$=1 (right).}
\label{fig2}
\end{figure}

{\it Current, charge and spectral density evolution:} 
We first analyze the transient regime for these basic quantities. 
Results for the current evolution are shown in the upper panel of Fig. \ref{fig2} for different values of the tunneling rates, phase difference $\phi/\pi=0.64$, and
for an initial condition $(n_{\uparrow}(0),n_{\downarrow}(0))=(0,1)$. 
We concentrate here in the highly transmitted, electron-hole symmetric case  
(i.e. $\epsilon_0 = 0$, $\Gamma_L \simeq \Gamma_R$) where the non-equilibrium effects that we discuss in this work
are more pronounced. Moreover, when $\Gamma \gg \Delta$ this case corresponds to a highly transmitted single-channel SAC.
In the electron-hole symmetric case the stationary ABSs are roughly located at
$\epsilon_A \simeq \tilde{\Delta} \sqrt{1 - \tau\sin^2(\phi/2)}$, where 
$\tau=4\Gamma_L\Gamma_R/\Gamma^2$ is
the normal transmission and $\tilde{\Delta} < \Delta$ varies from $\tilde{\Delta}
\sim \Gamma$ for $\Gamma \ll \Delta$ to $\tilde{\Delta} \rightarrow \Delta$
for $\Gamma \gg \Delta$ \cite{vecinoPRL}.  
As can be observed, the current reaches an asymptotic value smaller than the thermal equilibrium
stationary one (indicated by the arrows in Fig. \ref{fig2}), becoming even of 
opposite sign in the case of $\Gamma\lesssim\Delta$.

Further insight on this behavior can be obtained by analyzing the evolution of the central region occupied spectral
density (the method used to extract this quantity is described in the SM \cite{supplementary}). The lower panels in Fig. \ref{fig2} clearly illustrate
the process of formation of the subgap states, whose spectral weight mainly originates from the lower continuous spectrum at $\omega <-\Delta$. In addition,
the continuous spectrum exhibits oscillations which are gradually damped. As can 
be observed, it requires a characteristic time $\sim 2/|\epsilon_A|$ for the ABSs to become well defined \cite{comment-smooth}. The plots also show that while
for $\Gamma \gg \Delta$ the lower ABS becomes more populated, 
there is an inversion in their 
population for $\Gamma \lesssim \Delta$.   

\begin{figure}
\includegraphics[width=1\linewidth]{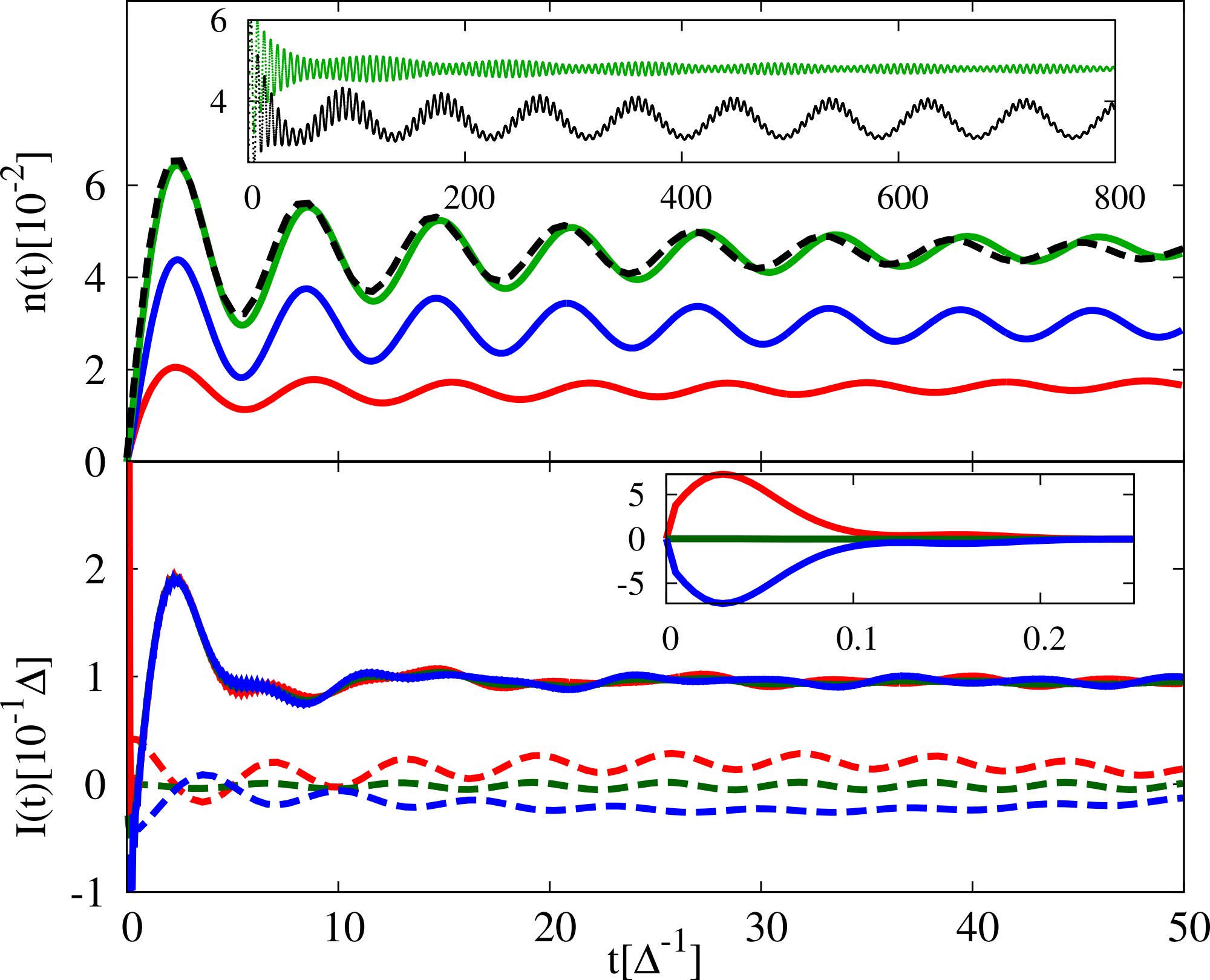}
\caption{(Color online):  The upper panel shows the ABSs populations $n_+$ (blue) and $n_-$ (red) for the case of the initial condition $(n_{\uparrow},n_{\downarrow})
= (0,1)$
 with $\Gamma=0.05\Delta$ and $\phi/\pi=0.64$ as determined from Eq. (\ref{analytic}). 
The dashed line corresponds to the total spin-up charge obtained numerically and the green line corresponds to $n_++n_-$. The inset in the upper panel illustrates the 
evolution of the level charge on a larger time scale in an electron-hole symmetric
($\epsilon_0=0$, green) and non-symmetric situation ($\epsilon_0=0.025\Gamma$, black). The black curve has
been shifted down for clarity. The lower panel illustrates the dependence of the 
transient current on the initial condition for the perfect transmission
case with $\Gamma/\Delta=10$ (full lines) and
$\Gamma/\Delta=0.05$ (dashed lines). The three initial conditions considered were
$(n_{\uparrow},n_{\downarrow})$=$(0,0)$ (red), $(0,1)$ (green) and 
$(1,1)$ (blue). The inset corresponds to a zoom at very short
times for the large $\Gamma$ case.}
\label{fig3}
\end{figure}

{\it Low tunneling rate regime:} In this regime the contribution from the 
continuum states to the level charge becomes negligible. As described in the SM \cite{supplementary}, this allows us to obtain an analytical expression for the population of the ABSs in this limit. Assuming the initial condition $(n_{\uparrow},
n_{\downarrow})=(0,1)$, these are given by 

\begin{equation}
 n_{\pm}(t)=\frac{\Gamma}{\pi}\int_{-W}^{-\Delta}\frac{(\omega \mp \Delta \cdot\epsilon_A/\Gamma)\left[\cos\left(\omega_{\pm}t\right)-1\right]}{
 \omega_{\pm}^2\sqrt{\omega^2-\Delta^2}}
 d\omega,
 \label{analytic}
\end{equation}
where $+/-$ corresponds to the upper/lower ABS
and $\omega_{\pm} = \omega \mp \epsilon_A$. As shown in upper panel of Fig. \ref{fig3}, the comparison of $n_++n_-$ with the total spin-up charge obtained
numerically for $\Gamma=0.05\Delta$ yields very good agreement. Both $n_+$ and $n_-$ exhibit an initial linear increase \cite{note}, with
a slope set by $\Gamma$, followed
by an oscillatory behavior with a characteristic period set by $\sim 2\pi/(\Delta\pm\epsilon_A)$. For $\Gamma \ll \Delta$ and $t >1/\Delta$, $n_{\pm}(t)$ 
is well described by the expression

\begin{equation}
n_{\pm}(t) = n_{\pm}(\infty) + \frac{\Gamma}{\Delta}\left(1\pm\frac{\epsilon_A}{\Gamma}\right)\frac{\sin\left((\Delta\pm\epsilon_A)t-\frac{\pi}{4}\right)}
{\sqrt{2\pi\Delta \cdot t}},
\label{analytic2}
\end{equation}
where $n_{\pm}(\infty)=\left(\Gamma/2 \pm \epsilon_A/\pi\right)/\Delta$. This expression indicates that the oscillatory behavior dies out 
with a weak power law (i.e., as $t^{-1/2}$). 
While in this limit the upper level is more populated than
the lower one, Eq. (\ref{analytic}) would predict an inversion of the relative populations for $\Gamma/\Delta \sim 1$. Although for 
$\Gamma\gtrsim \Delta$ 
the contribution from the continuum to the system dynamics can no longer be neglected, the prediction of the
population inversion is consistent with the numerical results for the current 
shown in the upper panel of Fig. \ref{fig2}. 

It is important to notice that when electron-hole symmetry is broken 
($\epsilon_0\neq0$) the switch-on process couples the two 
ABSs, thus generating an additional contribution to the level charge which oscillates with a frequency $2\epsilon_A$
(see inset in the upper panel Fig. \ref{fig3}). 
A more detailed analysis of this case is provided in the SM \cite{supplementary}. 

{\it Dependence on initial conditions:} The fact that the system reaches a metastable state suggests that this can be extremely sensitive to the initial 
conditions. This is true for the low $\Gamma$ regime, but this sensitivity gradually disappears for increasing $\Gamma$. This is illustrated in the lower panel of
Fig. \ref{fig3}, where the current in the perfect transmission case
for $\Gamma=0.05 \Delta$ and 
$\Gamma=10\Delta$ is shown for three different initial conditions 
$(n_{\uparrow},n_{\downarrow})$ =
$(0,0)$, $(0,1)$, $(1,1)$. It is observed that a positive 
or negative peak in $I$ of size $\sim \Gamma_L$ appears at
short times when the initial charge deviates from the expected stationary value. 
This peak, however, relaxes in a time scale set by $1/\Gamma$,
as illustrated in the inset of the lower panel of Fig. \ref{fig3}. 

\begin{figure}
\includegraphics[width=1\linewidth]{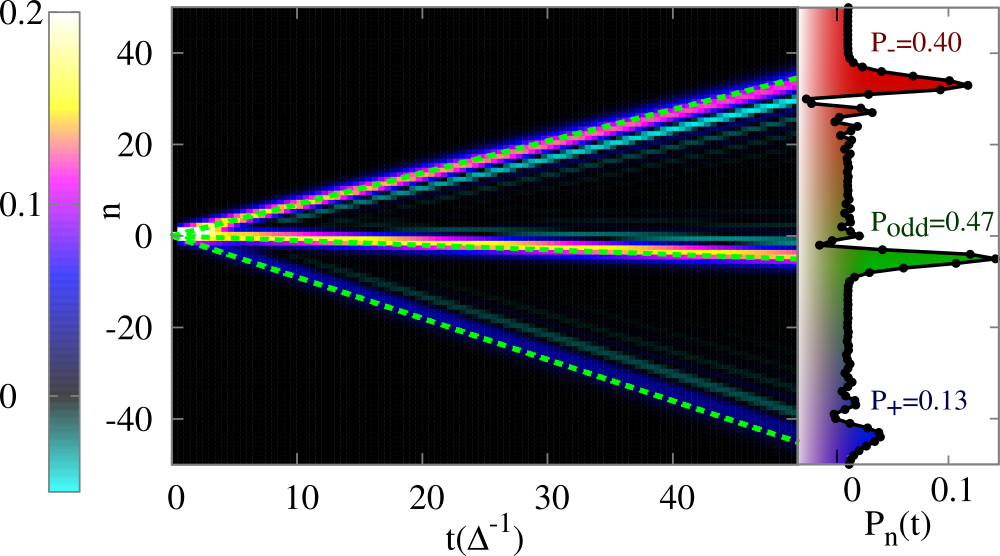}
\caption{(Color online): Density plot of $P_{n}(t)$ in the $(t,n)$ plane for the perfect transmission case with $\Gamma/\Delta=10$
and $\phi/\pi=0.64$. The dashed lines
are slopes corresponding to the current in the different coexisting many-body states (see text). The right insert shows a cut of $P_n(t)$ at $t=50 \Delta^{-1}$,
demonstrating that three asymptotic populations can be clearly identified.}
\label{fig4}
\end{figure}

{\it FCS analysis:} New light on the system dynamics can be shed by analyzing the quasi-probabilities $P_n(t)$. We first focus in the large $\Gamma/\Delta$
limit where, as commented before, the influence of the initial conditions is less pronounced. A clear picture emerges when $P_n(t)$ is analyzed as a function
of $n$ as in Fig. \ref{fig4}, where a density plot
of $P_n(t)$ on the $(t,n)$ plane is shown. One can clearly identify here three main lines with slopes $I_-$, $I_{odd}$ and $I_+$, which can be associated with
three different coexisting many-body states. The character of these states can be inferred from the slope values. Thus, $I_- \sim 2\partial\epsilon_A/\partial \phi$
corresponds to the system ground state; $I_+ \sim -I_-$ can be associated with the even excited state and $I_{odd} \sim 0$ would correspond to an odd state with a trapped quasiparticle within the ABSs (this last state is spin degenerate) \cite{zigrist}. A slight deviation of the slopes from these values arises from the contribution of the continuum to the supercurrent which becomes negligible in the 
$\Gamma/\Delta \rightarrow \infty$ limit. 

\begin{figure}
\includegraphics[width=1\linewidth]{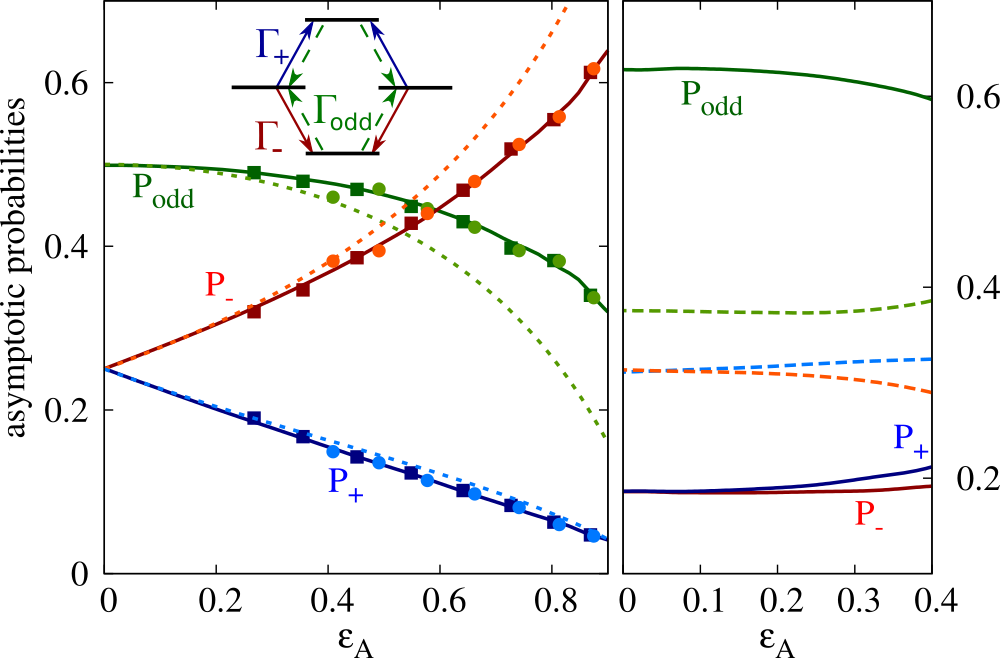}
\caption{(Color online): The left panel shows the asymptotic probabilities $P_{\pm}$ and $P_{odd}$ for $\Gamma/\Delta=10$ as a function of the ABS energy $\epsilon_A$. 
Full lines correspond to perfect transmission, squares to $\tau=0.95$ and dots to $\tau=0.9$. The dashed lines 
are results obtained with the simple rate equations model schematically depicted in the inset and assuming $\Gamma_{odd} \sim 1/\Delta$ and $\Gamma_{\pm} \sim 1/(\Delta\pm\epsilon_A)$.
The right panel corresponds to $\Gamma/\Delta=1$, perfect transmission and two different initial conditions $(n_{\uparrow},n_{\downarrow})$ = (0,1) (full lines) and (0,0) (dashed lines).}
\label{fig5}
\end{figure}

We can further characterize the metastable state by the weights $P_-$, $P_+$ and $P_{odd}$ of these three many-body states. An interesting feature of 
the system evolution is that once the ABSs become well defined (at times of the order of $2/|\epsilon_A|$) these weights remain nearly constant
as far as no external relaxation mechanism is operative on this time scale. 
In order to extract these weights two procedures can be used: i) directly
from the $P_n(t)$ at sufficiently large times, integrating the $P_n$ around the three peaks (see right insert in Fig. \ref{fig4});
ii) using the mean current, the mean noise and the normalization condition to set a system of three equations with three unknowns 
from which $P_+$, $P_-$ and $P_{odd}$ can be extracted (see SM \cite{supplementary}). Both procedures yield results which are in good agreement. 
Fig. \ref{fig5} shows the resulting asymptotic weights as a function of $\epsilon_A$ for two different
values of $\Gamma/\Delta$. In the case $\Gamma/\Delta=10$ (left panel) 
the odd state exhibits an increasing population with decreasing $\epsilon_A$ 
reaching a value of the order of 0.5 when 
$\epsilon_A \rightarrow 0$. At the same time $P_{\pm}$ tend to converge to the value 0.25 in this limit.  As in the case of the current the results for this large 
$\Gamma$ case are rather insensitive to the particular choice of the initial conditions. Moreover, the results in this limit are {\it universal} depending only
on $\epsilon_A$ irrespective of the junction transmission, as shown by squares
($\tau=0.95$) and circles ($\tau=0.9$) symbols in the left panel of Fig. \ref{fig5}.
These results are in remarkable agreement with those of Ref. \cite{landry} which
were obtained for a SAC with $\tau \sim 0.99$ and could be qualitatively understood in terms of a simple rate equation picture where the even ground state, the two odd states and the excited even state are connected by some effective rates, as depicted in the inset of Fig. \ref{fig5}.
Within this simplified picture and taking $\Gamma_{odd} \sim 1/\Delta$ and
$\Gamma_{\pm} \sim 1/(\Delta\pm\epsilon_A)$, i.e. inversely proportional to the 
energy distance between the states and the continuum, one obtains the results
indicated by the dashed lines in the left panel of Fig. \ref{fig5}, which are in 
good agreement with the numerical ones (see \cite{supplementary} for more details). 
In contrast, the results become increasingly sensitive to the initial conditions
for decreasing $\Gamma/\Delta$. As shown in the right panel of Fig. \ref{fig5}, 
for $\Gamma/\Delta = 1$ the asymptotic populations for different initial conditions strongly deviate from each other.

One should finally comment on the effect of additional relaxation mechanisms. In Ref. \cite{theory-poisoning} the relaxation through photon and phonon emission was analyzed
using a rate equation approach, obtaining a semiquantitative agreement with the experimental results of Ref. \cite{zigrist} in the stationary regime. This approach,
however, is not able to describe the initial stages of the ABSs formation which can be addressed by the present microscopic theory. In fact, one can identify 
$P_{odd}$ with the ``initial poisoning'' probability defined in Ref. \cite{zigrist}. A direct experimental test of the universal behavior predicted for this quantity would require analyzing the response to large voltage pulses. 

{\it Conclusions:} We have shown that the transient dynamics in the formation of a phase-biased superconducting nanojunction leads to a metastable state 
characterized by a non-equilibrium population of the system ABSs. Although in the quantum dot regime this state is strongly sensitive to the initial conditions, in the quantum point contact limit this sensitivity disappears and a universal asymptotic population is reached, only dependent on the
ABS energy levels. These findings shed light on the available experimental results like those of Refs. \cite{landry,zigrist} and could
be tested more thoroughly in future experiments.

We acknowledge discussions with C. Urbina and L. Arrachea
and financial support by Spanish MINECO through grant FIS2014-55486-P and the 
``Mar\'{\i}a de Maeztu'' Programme for Units of Excellence in R\&D (MDM-2014-0377). 
Computer resources by the Supercomputing and Visualization
Center of Madrid (CeSViMa) and the Spanish Supercomputing Network (RES) are also acknowledged.

\newpage

\widetext
\begin{appendices}
\section{\large{Supplementary Material on ``Andreev bound states formation and quasiparticle trapping in quench dynamics revealed by time-dependent counting statistics''}}

\section{Numerical evaluation of the Fredholm determinant}

In Refs. \cite{mukamel,chinos,us} it has been shown that for a spinless single-channel normal junction the GF can be written as

\begin{equation}
Z(\chi,t) = \det \left[ G\tilde{G}^{-1} \right]= \det \left[ G \left( g_0^{-1} - \tilde{\Sigma}  \right) \right]
\end{equation}
where $\tilde{G}$ and $G$ denote the dot Keldysh Green functions,   
$g_0$ corresponds to the uncoupled dot case and 
$\tilde{\Sigma}$ are the self-energies due to the coupling to the leads.  
In the quantities $\tilde{G}$ and $\tilde{\Sigma}$ the $tilde$ indicates 
the inclusion of the counting field in the tunnel amplitudes. 

\begin{figure}[hb!]
     \begin{minipage}{0.5\linewidth}
      \includegraphics[width=1.0\textwidth]{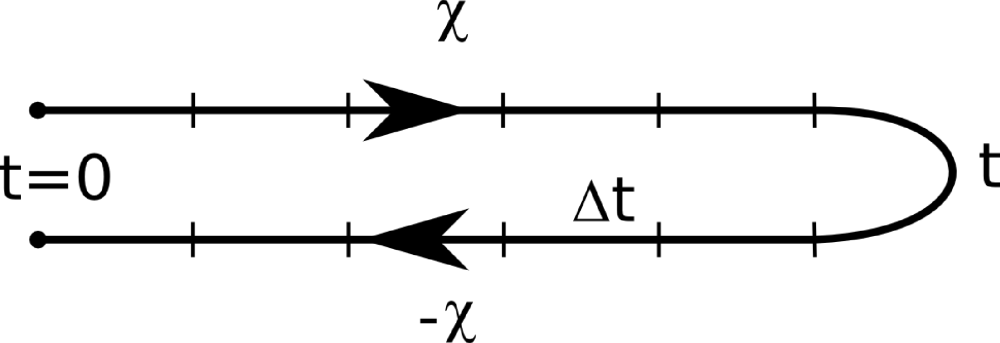}
     \end{minipage}
\caption{Keldysh contour considered to analyze the transient regime. $\chi$ indicates the counting field
changing sign on the two branches of the contour and $\Delta t$ corresponds to the time step in the
discretized calculation of the generating function $Z(\chi,t)$.}
\label{contour}
\end{figure}

As shown in \cite{kamenev} a simple discretized version of the inverse free dot level Green 
function on the Keldysh contour (as shown in Fig. \ref{contour}) is 

\begin{equation}
i g^{-1}_0 = \left(\begin{array}{cccc|cccc} -1 & & & & & & & -\rho \\
h_- & -1 & & & & & &  \\
& h_- & -1 & & & & &  \\
& & \ddots & \ddots & & & &  \\
\hline  
&  & & 1 & -1 & & &  \\
&  & & & h_+ & -1 & &   \\
&  &  &  & & \ddots & \ddots &  \\
&  &  & & & & h_+ & -1 \end{array} \right)_{2N\times2N} \; ,
\label{kamenev}
\end{equation}
where $h_{\pm} = 1 \mp i\epsilon_0 \Delta t$, $\Delta t$ indicates
the time step in the discretization with $N=t/\Delta t$. In this
expression $\rho$ determines the initial dot level charge $n(0)$ by $n(0) = \rho/(1+\rho)$.

In the superconducting case the above formalism has to be extended into the Nambu space, associating the electron components
with spin up states and hole components with spin down. This leads to the following expression for the GF

\begin{equation}
Z(\chi,t) = \det \left[ {\bf G} \left({\bf g}_0^{-1} - \tilde{\bf \Sigma}  \right) \right] \; ,
\label{GF-nambu}
\end{equation}
where the bold face letters indicate the Keldysh-Nambu extended quantities. Eq. (\ref{GF-nambu}) reduces to the expression (2) in the main text
by identifying ${\bf G}(\chi)^{-1}  \equiv \left({\bf g}_0^{-1} - \tilde{\bf \Sigma}  \right) $. For the uncoupled dot Green function we have

\begin{equation}
{\bf g}_0 = \left( \begin{array}{cc} g_{0,e} & 0 \\ 0 & g_{0,h} \end{array} \right), 
\end{equation}
where $g_{0,e}=g_0$ and $g_{0,h}(\epsilon_0,n_{\downarrow}(0)) = g_0(-\epsilon_0,(1-n_{\downarrow}(0)))$.

On the other hand, the lead self-energies would be given by
\begin{equation}
\tilde{\bf \Sigma}_{\nu} = \left( \begin{array}{cc} \tilde{\Sigma}_{\nu,11} & \tilde{\Sigma}_{\nu,12} \\ \tilde{\Sigma}_{\nu,21} & \tilde{\Sigma}_{\nu,22} \end{array} \right), 
\end{equation}
with

\begin{equation}
\tilde{\Sigma}^{\alpha\beta}_{\nu,jk}(t,t') = \theta(t)\theta(t') s_{\alpha}s_{\beta}s_j s_k \left(V^0_{\nu}\right)^2 e^{i(s_j-s_k)\phi_{\nu}/2} e^{i(s_{\alpha}s_j-s_{\beta}s_k)\chi_{\nu}/2} g^{\alpha\beta}_{jk}(t-t'),
\end{equation}
where $\alpha\beta \equiv \pm$ are the Keldysh indexes, $\nu=L,R$ denote the leads and
$s_{\pm}=\pm 1$ and $s_{j}= (-1)^{(j+1)}$. The uncoupled leads Green functions $g^{\alpha\beta}_{jk}(t-t')$ can be obtained by
Fourier transformation of the BCS Green functions in frequency representation \cite{Review}, leading to

\begin{equation}
g^{+-}_{11}(\tau) = g^{+-}_{22}(\tau) = -\frac{i\Delta}{\pi W} H^{(1)}_1(\tau\Delta) + 2\frac{e^{i\tau W}}{\pi^2W\tau}, 
\end{equation}
while $g^{-+}_{11}(\tau)=g^{-+}_{22}(\tau)=g^{+-}_{11}(\tau)^*$, and the anomalous components are 

\begin{equation}
g^{+-}_{12}(\tau) = g^{+-}_{21}(\tau) = -\frac{\Delta}{\pi W}H^{(1)}_0(\Delta\tau) - 2\frac{i\Delta}{\pi^2W}E_1(-i\Delta|\tau|)
\end{equation}
while $g^{-+}_{12}(\tau) = g^{-+}_{21}(\tau) = -g^{+-}_{12}(\tau)^*$. In the above expressions $H^{(1)}_{0,1}$ denote Hankel functions of the first kind, and
$E_1$ is the exponential integral with an imaginary argument. The other Keldysh components are obtained from the previous ones using the usual relations
in Keldysh space, i.e.

\begin{eqnarray}
g^{++}_{jk}(\tau) &=& \theta(\tau) g^{-+}_{jk}(\tau) + \theta(-\tau) g^{+-}_{jk}(\tau) \nonumber \\
g^{--}_{jk}(\tau) &=& \theta(-\tau) g^{-+}_{jk}(\tau)+ \theta(\tau) g^{+-}_{jk}(\tau),
\end{eqnarray}
with $j,k=1,2$. The time discretization of these expressions with a time step $\Delta t$ is straightforward. In order to get a stable numerical evaluation of the resulting
Fredholm determinant it is necessary to take $\Delta t \lesssim 1/W$. In addition we have found that the most stable numerical algorithm requires to take
the equal times self-energy component $\Sigma^{++,--}_{\nu,jk}(t,t)$ as $-(\Sigma^{+-}_{\nu,jk}(t,t)+\Sigma^{-+}_{\nu,jk}(t,t))/2$.

\begin{figure}[ht!]
     \begin{minipage}{0.5\linewidth}
      \includegraphics[width=1.0\textwidth]{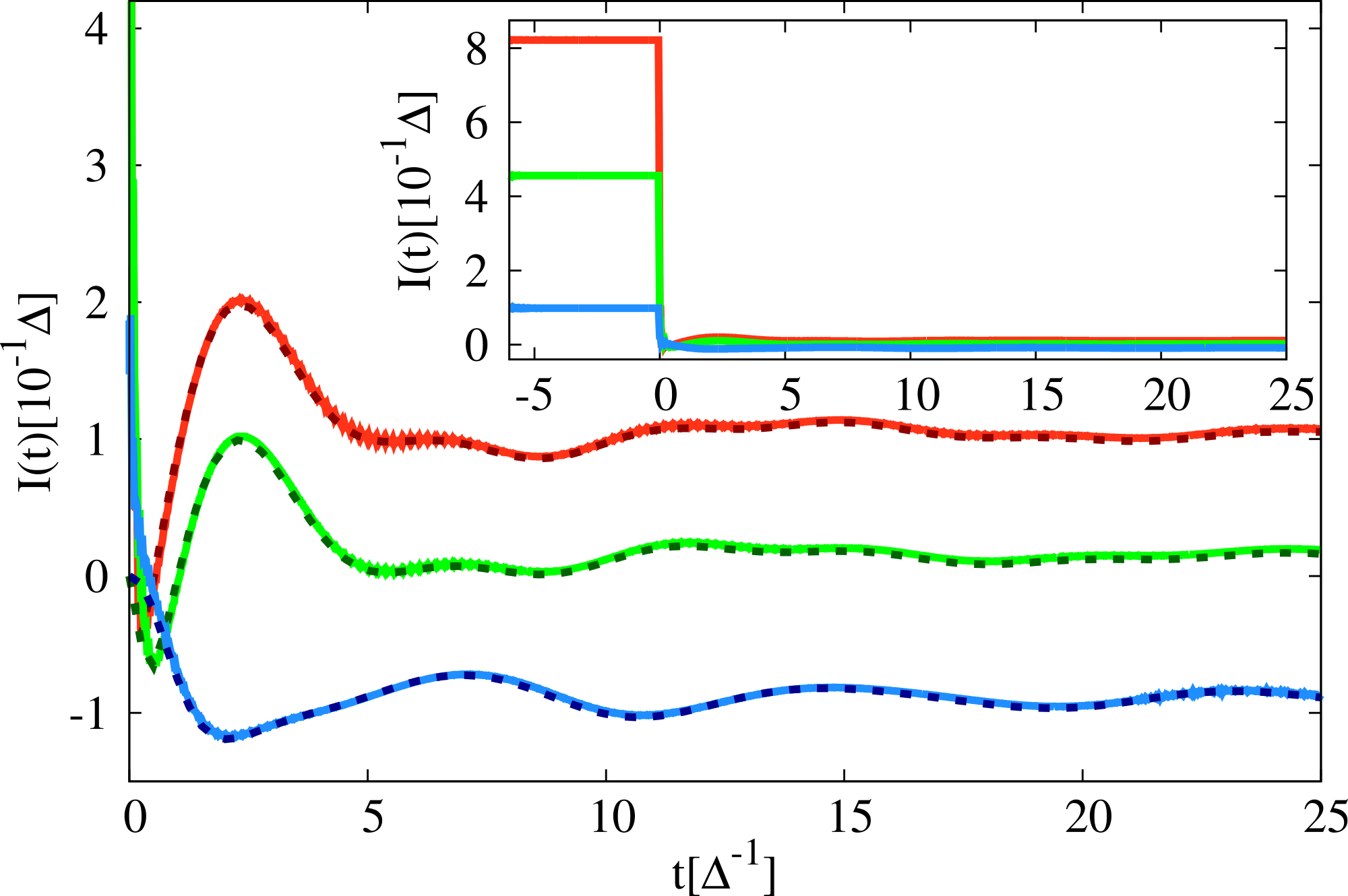}
     \end{minipage}
\caption{Current evolution for a large voltage step switch-off $eV_0/\Delta \sim 100$ for different $\Gamma$ values: 10 (red), 5 (green) and 1 (blue). The final phase
difference in all cases is $\phi/\pi=0.64$. The dashed lines correspond to the
abrupt connection to the leads case (same results as in Fig. 2 of the main text). 
The inset
shows the same results in a larger time and current scales. Notice that the expected ac Josephson oscillations in the current at $t<0$ are negligible in this scale.}
\label{voltage-switchoff}
\end{figure}

\section{Equivalence with initialization through a voltage switch off}

The same formalism described above can be extended to a voltage biased case, by
introducing a time-dependent phase $\phi(t)$. The bias voltage in this case is given
by $V(t) =\Phi_0 d\phi(t)/dt$, where $\Phi_0=\hbar/2e$. One can consider an initially coupled dot where 
a voltage step of the form $V(t) = V_0 \theta(- t)$ is applied. We have found that for $V_0 \gg \Gamma$ the
transient dynamics for the system becomes equivalent to the one obtained by
the sudden connection to the leads. This is illustrated in Fig. \ref{voltage-switchoff},
where the current evolution for a large voltage step $eV_0/\Delta \sim 100$ is
shown for different $\Gamma$ values (full lines) and compared with the
corresponding case for the abrupt coupling to the leads (dashed lines).
The reason for the equivalence is rather simple: due to the fast oscillating phase factors of the form $e^{iV_0t/2}$ entering in the self-energies $\tilde{\bold \Sigma}(t,t')$ connecting positive and negative times, the initial correlations between the
leads and the central region become negligible when $V_0 \gg \Gamma$.

\section{Evaluation of time-dependent spectral densities}

For the evaluation of the time-dependent occupied spectral densities we use the {\it auxiliary current method} as described in \cite{millis}. In this method
the dot is weakly coupled to an empty third electrode with a $\delta$-function spectral density, leading to a tunneling rate of the form
 $\Gamma_{A}^{\omega'}(w)=\eta\delta(\omega-\omega')$. The resulting current is directly related to the time-dependent
occupied spectral density $A(\omega,t)=1/(2\pi)\mbox{Im}[G^{+-}_{11}(\omega,t)-G^{-+}_{22}(\omega,t)]$, according to  
\begin{equation}
 A(\omega,t)=\frac{2h}{e\pi}\lim_{\eta\to0}\frac{I^{\omega}_A(t)}{\eta}
\end{equation}
where $I^{\omega}_A(t)$ is the current through the third lead.

\section{Analytic approach to transient mean charge and current}

When addressing the mean charge or the mean current, the problem can be
simplified by using the Keldysh formalism in the triangular form, where only the
retarded, advanced and Keldysh $+-$ components are involved in the Dyson
equations.  The retarded/advanced Green functions can be obtained by
the corresponding Dyson equations. For the case of an abrupt connection to
the leads these equations can be solved, leading to 

\begin{equation}
\hat{G}^{R,A}(t,t') = \theta(t)\theta(t') \hat{G}_{est}^{R,A}(t-t')
\end{equation}
where $\hat{G}_{est}^{R,A}(t-t')$ denotes the level stationary Green function
in Nambu space, whose Fourier transform has the simple form

\begin{equation}
\hat{G}_{est}^{R,A}(\omega) = \left(\begin{array}{cc} \omega - \epsilon_0 - \Gamma g_{11}^{R,A} & (\Gamma_L e^{i\phi_L}+\Gamma_R e^{i\phi_R}) g_{12}^{R,A} \\
 (\Gamma_L e^{-i\phi_L}+\Gamma_R e^{-i\phi_R}) g_{12}^{R,A} & \omega + \epsilon_0 - \Gamma g_{22}^{R,A}\end{array}\right)^{-1}
\end{equation}

The poles of $\hat{G}_{est}^{R,A}(\omega)$ correspond to the ABS at
$\pm \epsilon_A(\phi)$. In the limit $\Gamma/\Delta \ll 1$ the contribution from the
continuum spectrum for $|\omega|>\Delta$ becomes negligible and we can
approximate $\hat{G}^{R,A}(t,t')$ by

\begin{equation}
\hat{G}^{R,A}(t,t') \simeq \theta(t)\theta(t') \sum_{\pm} \left(\begin{array}{cc}
p_{\pm} & \pm p_{12} \\
\pm p^*_{12} & p_{\pm} \end{array}\right)e^{\pm i\epsilon_A(t-t')}
\label{retarded}
\end{equation}
where $|p_{12}|=\sqrt{p_+p_-}$. The ABS energies and the weights $p_{\pm}$ adopt a simple form when
$\epsilon_0 < \Gamma$, i.e. $\epsilon_A = \sqrt{\epsilon_0^2 + \Gamma^2 \cos^2(\phi/2) + (\Gamma_L-\Gamma_R)^2\sin^2(\phi/2)}$
and $p_{\pm} = (1\mp\epsilon_0/\epsilon_A)/2$.

The time-dependent level charge can then be obtained through the Dyson
equation for the Keldysh Green function $\hat{G}^{+-}$, which
for the case of an initially condition $(n_{\uparrow},n_{\downarrow})=(0,1)$
is given by

\begin{equation}
\hat{G}^{+-}(t,t') = \int dt_1 dt_2 \hat{G}^R(t,t_1)\hat{\Sigma}^{+-}(t_1,t_2)
\hat{G}^A(t_2,t')
\label{keldysh} 
\end{equation}
where $\hat{\Sigma}^{+-}$ is the Keldysh component of the self-energy 
coupling the level to the leads in Nambu space. 
Substitution of Eq. (\ref{retarded}) into Eq. (\ref{keldysh}) yields
an approximated $n_{\uparrow}(t)=-iG_{11}^{+-}(t,t)$, given by

\begin{eqnarray}
n_{\uparrow}(t) &= & \sum_{\pm} n_{\pm}(t) + n_{+-} \nonumber \\
n_{\pm}(t) &=& -\frac{2 p_{\pm}}{\pi}\int_{-\infty}^{-\Delta} \frac{\omega\Gamma\mp 4p_+p_- \Delta\cdot\epsilon_A}{\sqrt{\omega^2-\Delta^2} \omega_{\mp}}\left[1-\cos\left(\omega_{\mp}t\right)\right]d\omega \nonumber\\
n_{+-} &=&  \frac{2(p_+-p_-)}{\pi}\int_{-\infty}^{-\Delta} \frac{4p_+p_-\Delta\cdot\epsilon_A}{\omega_+\omega_-\sqrt{\omega^2-\Delta^2}}\left[1 + 
\cos\left(2\epsilon_At\right)-\cos(\omega_+t) - \cos(\omega_-t)\right] d\omega
\end{eqnarray}
where $\omega_{\pm} = \omega\pm\epsilon_A$. Similar expressions can be
derived for $n_{\downarrow} = -iG_{22}^{-+}(t,t)$.
As can be observed, the dot occupation is composed by contributions from the upper and lower ABS plus an 
interference term $n_{+-}$. Remarkably, this interference term, which indicates the entanglement between ABSs generated by the abrupt switch-on process,
vanishes in the electron-hole symmetric case (i.e. $\epsilon_0=0$). The asymptotic behavior of $n_{\pm}(t)$ has already been discussed in the main text for
this case and remains qualitatively the same for $\epsilon_0\ne0$. On the other hand, the interference term exhibits undamped oscillations of
frequency $2\epsilon_A$ and is  well approximated at times $t >1/\Delta$ as

\begin{equation}
n_{+-}(t) \simeq -\frac{\epsilon_0}{\pi\Delta}\left(1-\frac{\epsilon_0^2}{\epsilon_A^2}\right)\left[1+\cos(2\epsilon_At)+\sqrt{\frac{2\pi}{\Delta\cdot t}}
\cos(\epsilon_At)\sin\left(\Delta\cdot t-\frac{\pi}{4}\right)\right]
\end{equation} 
As a general remark, it is interesting to notice that for an initial condition with 
broken spin symmetry (i.e. $(1,0)$ or $(0,1)$) the system gets trapped in a magnetic state unless $\Gamma  \gg \Delta$. This is the case, for instance, in the upper
panel of Fig. 3, where $n_{\uparrow}(t) \ll 1$ and $n_{\downarrow}(t) = 1 - n_{\uparrow}(t) \simeq 1$ for the $\epsilon_0=0$ situation. 

\section{Extracting the asymptotic probabilities of many-body states from current and noise}

As can be observed in the lower panel of Fig. 4 of the main text, the 
quasi-probabilities
$P_n$ tend to concentrate at sufficiently long times into three peaks, which
we associate with the even ground state $(-)$, the excited even state $(+)$  and the odd state $(odd)$.  Each of these peaks have a total weight denoted by $P_{\pm}$
and $P_{odd}$. An alternative way to determine these weights, different from the direct
integration around the peaks commented in the text, is described below. 

In the long time limit the mean transferred charge and its second cumulant
can be written as

\begin{eqnarray}
\left\langle\left\langle n \right\rangle\right\rangle &\simeq& t \left(P_- I_- + P_+ I_+ + P_{odd} I_{odd} \right) \nonumber\\
\left\langle\left\langle n^2 \right\rangle\right\rangle &\simeq& t^2 \left[\left(
P_-I_-^2 + P_+ I_+^2 + P_{odd} I_{odd}^2\right) - \left(P_-I_-+P_+I_++P_{odd}I_{odd}\right)^2\right]\;.
\end{eqnarray}
These two equations, together with the normalization condition, $P_++P_-+P_{odd}=1$ lead to the many body probabilities 
\begin{equation}
 P_{\pm}=\frac{S+2\left(I^2-I_{odd}^2\right)}{4I_{A}^2}-\frac{I-I_{odd}}{2I_A}\left(\pm1+\frac{2I_{odd}}{I_A}\right)\;,
\end{equation}
where we have defined the current contribution of the ABSs as $I_A=I_--I_{odd}$, the total current as 
$I=\partial/\partial t \left\langle\left\langle n \right\rangle\right\rangle$ and the shot noise
$S=\partial^2/\partial t^2 \left\langle\left\langle n^2 \right\rangle\right\rangle$. Finally, 
the probability of populating the odd state can be computed by direct substitution in any of the three
equations.

\section{Interpretation in terms of rate equations}

A simple interpretation of the numerical results for the asymptotic probabilities
$P_{\pm}(t)$ and $P_{odd}(t)=P_{odd,\uparrow}(t)+P_{odd,\downarrow}(t)$ can be obtained assuming that they are governed
by simple rate equations with time dependent rates. More precisely, these equations
are

\begin{eqnarray}
\frac{dP_-}{dt} &=& -2\Gamma_{odd}(t) P_{-}  + \Gamma_{-}(t) P_{odd} \nonumber\\
\frac{dP_{odd}}{dt} &=& 2\Gamma_{odd}(t) P_{-}  - \left[\Gamma_{-}(t)+\Gamma_{+}(t)\right] P_{odd} +2\Gamma_{odd}(t) P_{+} \nonumber\\
\frac{dP_+}{dt} &=& \Gamma_{+}(t) P_{odd}  -2\Gamma_{odd}(t) P_{+}\;,
\end{eqnarray}
where $\Gamma_{odd}(t)$ and $\Gamma_{\pm}(t)$ are the time-dependent rates for transitions 
between states. Although these quantities are not well defined, an estimate
based on perturbation theory would suggest that they should be inversely 
proportional to the energy distance from the lower gap edge to the corresponding
state. 
For $\Gamma/\Delta\gg1$ the rates can be considered as time independent 
and approximated as  $\Gamma_{odd}\approx\Gamma^2/\Delta$ and $\Gamma_{\pm}\approx\Gamma^2/(\Delta\pm\epsilon_A)$. Using these estimates one obtains
the results indicated by the dashed lines in the left panel of Fig. 5 of the main text.



\end{appendices}

\begin{thebibliography}{99}
\bibitem{devoret} M.H. Devoret and R.J. Schoelkopf, Science {\bf 339}, 1169 (2013).
\bibitem{hofstetter} L. Hofstetter, S. Csonka, J. Nyg{\aa}rd and 
C. Sch\"onenberger, Nature {\bf 461}, 960 (2009).
\bibitem{hermann} L.G. Herrmann, F. Portier, P. Roche, A. Levy Yeyati, 
	T. Kontos and C. Strunk, Phys. Rev. Lett. {\bf 104}, 026801 (2010). 
\bibitem{landry} C. Janvier, L. Tosi, L. Bretheau, \c{C}. \''O. Girit, M. Stern, P. Bertet, P. Joyez, D. Vion, D. Esteve, M. F. Goffman, H. Pothier, C. Urbina, Science {\bf 349}, 1199 (2015).
\bibitem{review-majorana} For reviews see J. Alicea, Rep. Prog. Phys. {\bf 75}, 076501 (2012); C.W.J. Beenakker, Rev. Mod. Phys. {\bf 87}, 1037 (2015).
\bibitem{previous-transient} Some previous work have addresed time-resolved Josephson transport in superconducting nanojunctions, see for instance,
E. Perfetto, G. Stefanucci and M. Cini, Phys. Rev. B {\bf 80}, 205408 (2009);
G. Stefanucci, E. Perfetto and M. Cini, Phys. Rev. B {\bf 81}, 115446 (2010); B. Tarasinski, D. Chevallier, J. A. Hutasoit, B. Baxevanis and C.W.J. Beenakker, Phys. Rev. B
{\bf 92}, 144306 (2015); and J. Weston and X. Waintal, Phys. Rev. B {\bf 93}, 134506 (2016); 
\bibitem{various-qp} J. M. Martinis, M. Ansmann and J. Aumentado, Phys.
Rev. Lett. \textbf{103}, 097002 (2009); G. Catelani, R. J. Schoelkopf, M. H. Devoret and L.I. Glazman, Phys. Rev. B \textbf{84}, 064517 (2011); D. Rist\`e, C.C. Bultink, M.J. Tiggelman, R.N. Schouten, K.W. Lehnert, and L. DiCarlo, Nature comm. {\bf 4}, 1913 (2013); E.M. Levenson-Falk, F. Kos, R. Vijay, L. Glazman and I. Siddiqi,
Phys. Rev. Lett. {\bf 112}, 047002 (2014). 
\bibitem{zigrist} M. Zgirski, L. Bretheau, Q. Le Masne, H. Pothier,D. Esteve and C. Urbina, Phys. Rev. Lett. \textbf{106}, 257003 (2011).
\bibitem{theory-poisoning} D. Olivares, A. Levy Yeyati, L. Bretheau, C. Girit, H. Pothier and C. Urbina, Phys. Rev. B. 89, 104504 (2014); A. Zazunov, A. Brunetti, A. Levy Yeyati and R. Egger, Phys. Rev. B. 90, 104508 (2014).
\bibitem{levitov} L.S. Levitov in {Quantum noise in Mesoscopic Physics}, Edited by
Yu. V. Nazarov (Kluwer Academic Press, New York 2002), p. 373.
\bibitem{belzig} W. Belzig and Yu.V. Nazarov, Phys. Rev. Lett. {\bf 87}, 197006 (2001).
\bibitem{ramer} A. Shelankov and J. Rammer, Europhys. Lett. {\bf 63}, 485 (2003).
\bibitem{clerk} A recent work points out the possibility of negative FCS arising in 
more general situations due to quantum interference; see P.P. Hofer and A.A. Clerk, Phys. Rev. Lett. {\bf 116}, 013603 (2016).
\bibitem{mukamel} M. Esposito, U. Harbola and S. Mukamel, Rev. Mod. Phys. {\bf 81}, 1665 (2009).
\bibitem{chinos} G.-Tang, F. Xu and J. Wang, Phys. Rev. B {\bf 89}, 205310 (2014). 
\bibitem{us} R. Seoane, R. Avriller, R.C. Monreal, A. Mart\'{\i}n-Rodero and
A. Levy Yeyati, Phys. Rev. B {\bf 92}, 125435 (2015).
\bibitem{supplementary} See Supplementary Material, which includes Refs. \cite{kamenev,chen}.
\bibitem{kamenev} A. Kamenev, {\it Field Theory of Non-Equilibrium Systems}, (Cambridge University, Cambridge 2011).
\bibitem{chen} H.T. Chen, G. Cohen, A.J. Millis and D.R. Reichman, Phys. Rev. B
{\bf 93}, 174309 (2016).
\bibitem{comment} This corresponds to a relatively large broadening. As a reference in Ref. \cite{zigrist} relaxation times of the
order of 1 to 100 $\mu s$ have been reported, which would correspond to inelastic rates $1/\tau_{in} \sim 10^{-5}-10^{-7} \Delta$ in the case of Al.
\bibitem{vecinoPRL} A. Levy Yeyati, A. Mart\'{\i}n-Rodero and E. Vecino Phys. Rev. Lett. {\bf 91}, 266802 (2003); A. Mart\'{\i}n-Rodero and A. Levy Yeyati, Adv. Phys. \textbf{60}, 899 (2011).
\bibitem{comment-smooth} We have further checked that the results are robust when the instantaneous switch on is replaced by a smooth one provided that its
rate is faster than the characteristic times for the ABS formation.
\bibitem{note} More strictly, Eq. (\ref{analytic}) indicates an initial increase of $n_{\pm}(t)$ as $t^2$ for $t\lesssim 1/W$ followed by a linear increase for $1/W<t<1/\Delta$. 
\end{thebibliography}

\begin{references} 
\bibitem{mukamel} M. Esposito, U. Harbola and S. Mukamel Rev. Mod. Phys. {\bf 81}, 1665 (2009).
\bibitem{chinos} G.-Tang, F. Xu and J. Wang, Phys. Rev. B {\bf 89}, 205310 (2014).
\bibitem{us} R. Seoane Souto, R. Avriller, R. C. Monreal, A. Mart\'{\i}n-Rodero, and A. Levy Yeyati Phys. Rev. B {\bf 92}, 125435 (2015).
\bibitem{kamenev} A. Kamenev, {\it Field Theory of Non-Equilibrium Systems} (Cambridge University, Cambridge, 2011).
\bibitem{Review}A. Mart\'{\i}n-Rodero and A. Levy Yeyati, Adv. Phys. \textbf{60}, 899 (2011).
\bibitem{millis} H. T. Chen, G. Cohen,A. J. Millis and D. R. Reichman Phys. Rev. B {\bf 93}, 174309 (2016)
\end{references}
\end{document}